\documentclass[showpacs,preprintnumbers,amsmath,amssymb]{revtex4}

\usepackage{graphicx}
\usepackage{bm}
\usepackage{color}

\newcommand{\bq}{\begin{equation}}
\newcommand{\eq}{\end{equation}}
\newcommand{\bqa}{\begin{eqnarray}}
\newcommand{\eqa}{\end{eqnarray}}
\newcommand{\nn}{\nonumber \\}

\def\be     {\begin{equation}}
\def\ee     {\end{equation}}
\def\bea        {\begin{eqnarray}}
\def\eea        {\end{eqnarray}}
\def\bnn    {\begin{eqnarray*}}
\def\enn    {\end{eqnarray*}}

\begin{document}

\title{Comparison between renormalization group derived emergent dual holography and string theory based holographic duality}
\author{Ki-Seok Kim}
\affiliation{Department of Physics, POSTECH, Pohang, Gyeongbuk 37673, Korea}

\date{\today}

\begin{abstract}
Recently, we derived an effective holographic dual field theory of the Einstein-Klein-Gordon type for correlated $O(N)$ spin models, which gives rise to coupled equations of emergent dynamical metric tensor and holographically dual scalar field in the large $N$ limit. This renormalization group (RG) derived holographic dual field theory manifests functional RG flows of both coupling functions and order-parameter fields through natural appearance of an extra-dimensional space. It turns out that the emergent holographic dual description looks similar to the string-theory based conventionally utilized dual-holographic theory. On the other hand, the diffeomorphism invariance is explicitly broken due to gauge fixing in the Wilsonian RG-transformation procedure, which has been performed rather approximately. To clarify the mathematical structure of the RG-derived holographic dual field theory, we rewrite the string-theory based conventionally utilized dual holographic effective field theory based on the ADM decomposition of the metric tensor. This comparison leads us to claim that the RG-derived emergent holographic dual field theory takes into account higher-derivative curvature terms with gauge fixing in the string-theory based conventionally utilized Einstein-Klein-Gordon theory, giving rise to the RG flow of the metric tensor beyond the AdS (anti-de Sitter space) geometry. Furthermore, we compare the Hamilton-Jacobi equation for the effective IR on-shell action of the string-theory based conventionally utilized dual holographic effective theory with that of the RG-based holographic dual field theory. It turns out that the effective IR on-shell action of the string-theory based dual holography can be identified with the IR boundary effective action of the RG-based emergent holographic dual description, where the Wilsonian RG-transformation procedure may be regarded as an inverse process of the holographic renormalization. This demonstration leads us to propose an effective dual holographic field theory with the diffeomorphism invariance and higher derivative curvature terms, where the IR boundary condition is newly introduced to clarify the deep connection between UV microscopic and IR macroscopic degrees of freedom.
\end{abstract}

%\pacs{71.10.Hf, 71.30.+h, 71.10.-w, 71.10.Fd}

\maketitle

\section{Introduction}

Since the holographic duality conjecture \cite{Holographic_Duality_I,Holographic_Duality_II,Holographic_Duality_III,Holographic_Duality_IV, Holographic_Duality_V,Holographic_Duality_VI,Holographic_Duality_VII} had been proposed more than twenty years before, this theoretical framework has been applied to various strongly correlated systems, which cover confinement $\&$ chiral symmetry breaking in quantum chromodynamics \cite{Holographic_Duality_II,Holographic_Duality_III} and superconductivity $\&$ non-Fermi liquids in condensed matter physics \cite{Nakayama:2013is,Superconductivity_Holography,Benincasa:2012wu,Erdmenger:2013dpa,OBannon:2015cqy,Erdmenger:2016jjg,NFL_Holography, FQHE_Holography,MIT_Holography,Graphene_AdS_Description}. Although this non-perturbative theoretical framework gives remarkable solutions sometimes in the view of universality, there still exists one essential unsatisfactory point: it is not clear how to relate UV microscopic degrees of freedom with IR emergent macroscopic observables. To overcome this difficulty, researchers have tried to derive an effective holographic dual field theory based on renormalization group (RG) transformations \cite{RG_Holography_I,RG_Holography_II,RG_Holography_III,RG_Holography_IV,RG_Holography_V,RG_Holography_VI,RG_Holography_VII,RG_Holography_VIII,
RG_Holography_IX,RG_Holography_X,RG_Holography_XI,RG_Holography_XII,RG_Holography_XIII,RG_Holography_XIV,RG_Holography_XV,RG_Holography_XVI,
RG_Holography_XVII,RG_Holography_XVIII,RG_Holography_XIX,RG_Holography_XX,SungSik_Holography_I,SungSik_Holography_II,SungSik_Holography_III,
Einstein_Klein_Gordon_RG_Kim,Einstein_Dirac_RG_Kim,RG_GR_Geometry_I_Kim,RG_GR_Geometry_II_Kim,Kondo_Holography_Kim,Kitaev_Entanglement_Entropy_Kim,
RG_Holography_First_Kim}.

Recently, we derived an effective field theory of the Einstein-Klein-Gordon type, which manifests functional RG flows of both coupling functions and order-parameter fields through natural appearance of an extra-dimensional space \cite{Einstein_Klein_Gordon_RG_Kim,Einstein_Dirac_RG_Kim,RG_GR_Geometry_I_Kim,RG_GR_Geometry_II_Kim}. These functional RG flows are described by coupled equations of emergent dynamical metric tensor (coupling function) and holographically dual scalar field (order parameter), respectively, in the large $N$ limit, where $N$ is the flavor degeneracy of original matter fields. Such non-perturbatively renormalized metric and dual order-parameter field appear in the IR boundary action of original matter fields, which play the role of vertex and self-energy corrections in the original UV field theory. This theoretical framework serves as an effective mean-field theory in the large $N$ limit, the IR boundary condition of which corresponds to a mean-field equation of the dual order-parameter field in the presence of self-consistently renormalized coupling functions.

Although this RG-based holographic dual field theory looks similar to the string-theory based conventionally-utilized dual holographic field theory, one important problem is that the Wilsonian RG-transformation procedure is implemented rather approximately, more precisely, in the linearized fashion. As a result, the functional RG flows of both the metric and dual scalar field are given by linearized Einstein scalar equations along the extra-dimensional space. In addition, the diffeomorphism invariance is explicitly broken due to gauge fixing in the Wilsonian RG-transformation procedure, where the lapse function and the shift vector field are set to be $1$ and $0$, respectively, in the ADM formulation of the metric tensor \cite{ADM_Hamiltonian_Formulation}. One may suspect that the breakdown of the diffeomorphism invariance is related with the approximate implementation of the RG-transformation procedure.

In the present study, we compare the RG-based emergent holographic dual field theory with the string-theory based conventional dual holographic theory carefully. To clarify the mathematical structure of the RG-derived holographic dual field theory, we rewrite the string-theory based conventionally utilized dual holographic effective field theory based on the ADM decomposition of the metric tensor \cite{ADM_Hamiltonian_Formulation}. This comparison leads us to claim that the RG-derived emergent holographic dual field theory takes into account higher-derivative curvature terms with gauge fixing in the string-theory based conventionally utilized Einstein-Klein-Gordon theory, giving rise to the RG flow of the metric tensor beyond the AdS (anti-de Sitter space) geometry. Furthermore, we compare the Hamilton-Jacobi equation for the effective IR on-shell action of the string-theory based dual holographic effective theory \cite{Holographic_Duality_IV,Holographic_Duality_V,Holographic_Duality_VI} with that of the RG-derived holographic dual field theory. It turns out that the effective IR on-shell action of the string-theory based dual holography can be identified with the IR boundary effective action of the RG-based emergent holographic dual description, where the Wilsonian RG-transformation procedure my be regarded as an inverse process of the holographic renormalization \cite{Holographic_Duality_IV,Holographic_Duality_V,Holographic_Duality_VI}. This demonstration leads us to propose an effective dual holographic field theory with the diffeomorphism invariance and higher derivative curvature terms, where the IR boundary condition is newly introduced to clarify the deep connection between UV microscopic and IR macroscopic degrees of freedom.

\section{Holographic dual effective field theory}

We start our discussions, reviewing our recent works on RG-based holographic dual effective field theories \cite{Einstein_Klein_Gordon_RG_Kim,Einstein_Dirac_RG_Kim,RG_GR_Geometry_I_Kim,RG_GR_Geometry_II_Kim}. We considered an effective field theory for interacting scalar fields,
\bqa && Z[\varphi_{ext}(x)] = \int D \phi_{\alpha}(x) D g_{B}^{\mu\nu}(x) ~ \delta\Big(g_{B}^{\mu\nu}(x) - \delta^{\mu\nu}\Big) ~ \exp\Big[ - \int d^{D} x \sqrt{g_{B}(x)} \Big\{ g_{B}^{\mu\nu}(x) [\partial_{\mu} \phi_{\alpha}(x)] [\partial_{\nu} \phi_{\alpha}(x)] \nn && + m^{2} \phi_{\alpha}^{2}(x) + \xi R_{B}(x) \phi_{\alpha}^{2}(x) + \frac{u}{2 N} \phi_{\alpha}^{2}(x) \phi_{\beta}^{2}(x) - i \varphi_{ext}(x) \phi_{\alpha}^{2}(x) \Big\} \Big] . \eqa
Here, $\phi_{\alpha}(x)$ is a scalar field on $D-$dimensional Euclidean spacetime, where the vector index $\alpha$ runs from $1$ to $N$. $N = 1$, $N = 2$, and $N = 3$ correspond to Ising, XY, and Heisenberg model for correlated spins, respectively \cite{Large_N_phi4_Theory}. We put this effective field theory on a curved spacetime formally \cite{Coupling_Scalarfields_Riccicurvature}, where the background metric $g_{B}^{\mu\nu}(x)$ is Euclidean $\delta^{\mu\nu}$. $m^{2}$ is the mass parameter, tuned to cause a quantum phase transition from a disordered paramagnetic state to an ordered antiferromagnetic phase. $u$ is the strength of effective interactions of the $O(N)$ singlet $\phi_{\alpha}^{2}(x)$, which results in an interacting fixed point for quantum criticality. $\xi R_{B}(x) \phi_{\alpha}^{2}(x)$ is a curvature induced mass term \cite{Coupling_Scalarfields_Riccicurvature}, where $R_{B}(x)$ is the Ricci scalar, which vanishes at the UV boundary. To measure correlation functions of the $O(N)$ singlet, we introduced an external potential field $\varphi_{ext}(x)$.

Recently, we performed Wilsonian RG transformations to this effective field theory \cite{Einstein_Klein_Gordon_RG_Kim,Einstein_Dirac_RG_Kim,RG_GR_Geometry_I_Kim,RG_GR_Geometry_II_Kim}. The only nontrivial aspect in this RG-transformation procedure is that we implemented such transformations in the presence of background fields. We introduced an order-parameter field $\varphi^{(0)}(x)$ dual to the $O(N)$ singlet $\phi_{\alpha}^{2}(x)$, taking the Hubbard-Stratonovich transformation in the interaction term $\frac{u}{2 N} \phi_{\alpha}^{2}(x) \phi_{\beta}^{2}(x)$. Then, we separated all dynamical fields of $\phi_{\alpha}(x)$ and $\varphi^{(0)}(x)$ to their slow and fast modes, and performed the path integral with respect to all the fast degrees of freedom. The path integral for quantum fluctuations of heavy dual scalar fields $\varphi^{(0)}(x)$ results in an effective interaction term $\phi_{\alpha}^{2}(x) \phi_{\beta}^{2}(x)$, given by essentially the same expression as the above. To deal with this newly generated interaction term, we performed the Hubbard-Stratonovich transformation once again and introduced the second dual scalar field $\varphi^{(1)}(x)$ to the resulting effective action. The path integral for high-energy fluctuations of original matter fields $\phi_{\alpha}(x)$ gives rise to renormalization in the metric tensor, which corresponds to vertex corrections of coupling constants. Updating the bare metric $g^{\mu\nu (0)}(x) = g_{B}^{\mu\nu}(x)$ with the renormalized one $g^{\mu\nu (1)}(x)$, we took the same RG transformation with respect to the original matter field $\phi_{\alpha}(x)$ and the second dual order-parameter field $\varphi^{(1)}(x)$. Repeating these Wilsonian RG transformations in the presence of dual background fields, we could obtain \cite{Comment_Higher_Spin_GT}
\bqa && Z[\varphi_{ext}(x)] = \int D \phi_{\alpha}(x) D \varphi(x,z) D g_{\mu\nu}(x,z) ~ \delta\Big(g^{\mu\nu}(x,0) - \delta^{\mu\nu}\Big) ~ \delta\Big(\varphi(x,0) - \varphi_{ext}(x)\Big) \nn && \delta\Big\{\partial_{z} g^{\mu\nu}(x,z) - g^{\mu\gamma}(x,z) \Big(\partial_{\gamma} \partial_{\gamma'} G_{xx'}[g_{\mu\nu}(x,z),\varphi(x,z)]\Big)_{x' \rightarrow x} g^{\gamma'\nu}(x,z) \Big\} \nn && \exp\Big[ - \int d^{D} x \sqrt{g(x,z_{f})} \Big\{ g^{\mu\nu}(x,z_{f}) [\partial_{\mu} \phi_{\alpha}(x)] [\partial_{\nu} \phi_{\alpha}(x)] + [m^{2} - i \varphi(x,z_{f})] \phi_{\alpha}^{2}(x) + \xi R(x,z_{f}) \phi_{\alpha}^{2}(x) \Big\} \nn && - N \int_{0}^{z_{f}} d z \int d^{D} x \sqrt{g(x,z)} \Big\{ \frac{1}{2u} [\partial_{z} \varphi(x,z)]^{2} + \frac{\mathcal{C}_{\varphi}}{2} g^{\mu\nu}(x,z) [\partial_{\mu} \varphi(x,z)] [\partial_{\nu} \varphi(x,z)] + \mathcal{C}_{\xi} R(x,z) [\varphi(x,z)]^{2} \nn && + \mathcal{V}_{eff}[\varphi(x,z)] + \frac{1}{2 \kappa} \Big( R(x,z) - 2 \Lambda \Big) \Big\} \Big] . \eqa
Here, the extra-dimensional space $z$ results from the continuum approximation for the iteration number of RG transformations. The RG flow of the metric tensor, given in the $\delta-$function constraint, is realized by the high-energy propagator of original matter fields,
\bqa && \Big\{- \frac{1}{\sqrt{g(x,z)}} \partial_{\mu} \Big( \sqrt{g(x,z)} g^{\mu\nu}(x,z) \partial_{\nu} \Big) + \frac{1}{\epsilon} [m^{2} - i \varphi(x,z)] \Big\} G_{xx'}[g_{\mu\nu}(x,z),\varphi(x,z)] = \frac{1}{\sqrt{g(x,z)}} \delta^{(D)}(x-x') . \nn \eqa
$\epsilon \propto d z$ is the RG-transformation scale, where $f \epsilon$ with the iteration number $f$ of RG transformations covers the whole integration region of original matter fields.

We emphasize again that this RG-transformation procedure is nothing special except for the introduction of the background dual field variable. Both bare values of the metric tensor $g^{\mu\nu}(x,0) = \delta^{\mu\nu}$ and the dual scalar field $\varphi(x,0) = \varphi_{ext}(x)$, identified with the UV boundary condition for each dual field, RG-flow through the extra-dimensional space, described by the $\delta-$function constraint for the metric tensor and the bulk equation of motion for the dual scalar field. The RG flow of the metric tensor, $\partial_{z} g^{\mu\nu}(x,z) = g^{\mu\gamma}(x,z) \Big(\partial_{\gamma} \partial_{\gamma'} G_{xx'}[g_{\mu\nu}(x,z),\varphi(x,z)]\Big)_{x' \rightarrow x} g^{\gamma'\nu}(x,z)$, is given by quantum fluctuations of original matter fields as mentioned above and noticed from the Green's function with the RG scale $\epsilon$. We emphasize that non-locality in this RG flow of the metric tensor is inevitable, responsible for the appearance of higher spin fields, but neglected as the zeroth-order approximation \cite{Comment_Higher_Spin_GT}. Although the Einstein-Hilbert action $\frac{1}{2 \kappa} \Big( R(x,z) - 2 \Lambda \Big)$ results from quantum fluctuations of original matter fields, referred to as induced gravity \cite{Gradient_Expansion_Gravity_I,Gradient_Expansion_Gravity_II}, this metric tensor is not dynamical but rather ``static", where the kinetic energy along the extra-dimensional space does not appear. In this respect the Einstein-Hilbert action contributes to a vacuum energy. Below, we discuss the case when the metric tensor becomes dynamical. The $D-$dimensional kinetic energy of the order-parameter field, $\frac{\mathcal{C}_{\varphi}}{2} g^{\mu\nu}(x,z) [\partial_{\mu} \varphi(x,z)] [\partial_{\nu} \varphi(x,z)]$, with the curvature-induced effective mass term $\mathcal{C}_{\xi} R(x,z) [\varphi(x,z)]^{2}$ and the effective potential part $\mathcal{V}_{eff}[\varphi(x,z)]$ in the bulk action also originates from quantum fluctuations of original matter fields during the RG-transformation procedure. It is straightforward to obtain the equation of motion for the dual scalar field $\varphi(x,z)$, the RG flow of which is realized in a self-consistently and nonlinearly intertwined way with that of the metric tensor. Fully renormalized values of the metric tensor and the order-parameter field appear in the IR effective action for the original matter field, which play the role of renormalized coupling functions and self-energy corrections in the effective IR action, respectively. Minimizing the above effective action with respect to both $g^{\mu\nu}(x,z_{f})$ and $\varphi(x,z_{f})$, we obtain two coupled constraint equations, which specify the IR boundary conditions for both the metric tensor and order-parameter field, respectively. The IR boundary condition for the order-parameter field is nothing but the mean-field equation in the large $N$ limit, where non-perturbative renormalization effects of the coupling functions have been introduced with self-consistency.

The absence of dynamics in the metric tensor results from the lack of effective interactions of the energy-momentum tensor. Now, one may introduce effective tensor-type interactions as follows \cite{TTbar_Deformation}
\bqa && Z[\varphi_{ext}(x)] = \int D \phi_{\alpha}(x) D g_{B}^{\mu\nu}(x) ~ \delta\Big(g_{B}^{\mu\nu}(x) - \delta^{\mu\nu}\Big) ~ \exp\Big[ - \int d^{D} x \sqrt{g_{B}(x)} \Big\{ g_{B}^{\mu\nu}(x) [\partial_{\mu} \phi_{\alpha}(x)] [\partial_{\nu} \phi_{\alpha}(x)] \nn && + m^{2} \phi_{\alpha}^{2}(x) + \xi R_{B}(x) \phi_{\alpha}^{2}(x) + \frac{u}{2 N} \phi_{\alpha}^{2}(x) \phi_{\beta}^{2}(x) + \frac{\lambda}{2 N} T^{\mu\nu}(x) T_{\mu\nu}(x) - i \varphi_{ext}(x) \phi_{\alpha}^{2}(x) \Big\} \Big] , \eqa
where $T_{\mu\nu}(x)$ is the energy-momentum tensor current and $\lambda$ is the strength of their effective interactions. Implementing essentially the same and typical RG-transformation procedure with both Hubbard-Stratonovich fields of the metric tensor and the scalar order-parameter field, we found the following holographic dual effective field theory \cite{Einstein_Klein_Gordon_RG_Kim,Einstein_Dirac_RG_Kim}
\bqa && Z[\varphi_{ext}(x)] = \int D \phi_{\alpha}(x) D \varphi(x,z) D g_{\mu\nu}(x,z) ~ \delta\Big(g^{\mu\nu}(x,0) - \delta^{\mu\nu}\Big) \delta\Big(\varphi(x,0) - \varphi_{ext}(x)\Big) \nn && \exp\Bigg[ - \int d^{D} x \sqrt{g(x,z_{f})} \Big\{ g^{\mu\nu}(x,z_{f}) [\partial_{\mu} \phi_{\alpha}(x)] [\partial_{\nu} \phi_{\alpha}(x)] + [m^{2} - i \varphi(x,z_{f})] \phi_{\alpha}^{2}(x) + \xi R(x,z_{f}) \phi_{\alpha}^{2}(x) \Big\} \nn && - N \int_{0}^{z_{f}} d z \int d^{D} x \sqrt{g(x,z)} \Bigg\{ \frac{1}{2u} [\partial_{z} \varphi(x,z)]^{2} + \frac{\mathcal{C}_{\varphi}}{2} g^{\mu\nu}(x,z) [\partial_{\mu} \varphi(x,z)] [\partial_{\nu} \varphi(x,z)] + \mathcal{C}_{\xi} R(x,z) [\varphi(x,z)]^{2} \nn && + \mathcal{V}_{eff}[\varphi(x,z)] - \frac{1}{2 \lambda} \Big\{\partial_{z} g^{\mu\nu}(x,z) - g^{\mu\gamma}(x,z) \Big(\partial_{\gamma} \partial_{\gamma'} G_{xx'}[g_{\mu\nu}(x,z),\varphi(x,z)]\Big)_{x' \rightarrow x} g^{\gamma'\nu}(x,z)\Big\}^{2} \nn && + \frac{1}{2 \kappa} \Big( R(x,z) - 2 \Lambda \Big) \Bigg\} \Bigg] . \label{RG_Holography_EFT} \eqa
It is clear that not only the order-parameter field but also the metric tensor becomes dynamical. The large $N$ limit gives rise to coupled equations for both the metric tensor and the scalar field, describing their renormalization effects.

It is true that this RG-derived holographic dual effective field theory is rather similar to the conventional dual holographic formulation based on the string theory except for the appearance of the Green's function in the bulk dynamics of the metric tensor. Of course, the main difference between the two would be in the IR boundary condition, where only the RG-based dual holography reveals the clear connection between UV microscopic and IR macroscopic degrees of freedom. However, we have to point out that the Wilsonian RG-transformation procedure has been performed in the linearized fashion. Since the RG transformation was implemented with the introduction of dual field variables, the limit of this approximation is not clear at all. Previously, we applied essentially the same RG-transformation technique to the Kondo problem, and could describe the crossover behavior from the decoupled local-moment state to the local Fermi-liquid state without any artificial phase transitions \cite{Kondo_Holography_Kim}. Even if we did not introduce renormalization effects of the Kondo coupling constant but take into account the RG flow of the order-parameter field, both the specific heat and the spin susceptibility for the impurity dynamics shows reasonable match with the Bethe ansatz results. In this respect it is natural to expect that all-loop order quantum corrections are resumed to result in the form of an effective holographic dual field theory as an effective mean-field description in the large $N$ limit.

Still, there is an unsatisfactory point in this formulation. We recall the gauge fixing condition that $g^{DD}(x,z) = 1$ and $g^{\mu D}(x,z) = 0$, where $\mu$ spans the Euclidean spacetime dimension $0, ..., D-1$. In other words, the diffeomorphism invariance is broken explicitly in the RG-transformation procedure. Moreover, this gauge fixing prohibits us from considering the Ward identity \cite{Holographic_Ward_Identity} involved with the diffeomorphism invariance. This Ward identity can be used to justify the approximation in the RG-transformation procedure. In this respect it is meaningful to impose the diffeomorphism invariance to the RG-derived dual holographic effective field theory. Unfortunately, we do not know how to implement the RG transformation in the way of diffeomorphism invariance.

In this study, we approach this problem in a different angle, starting from the string-theory based conventionally utilized effective action,
\bqa && \mathcal{S}_{eff}^{GR} = N \int_{0}^{z_{f}} d z \int d^{D} x \sqrt{\mathcal{G}(x,z)} \Big\{ \frac{\mathcal{C}_{\varphi}}{2} \mathcal{G}^{MN}(x,z) [\partial_{M} \varphi(x,z)] [\partial_{N} \varphi(x,z)] + \mathcal{C}_{\xi} \mathcal{R}(x,z) [\varphi(x,z)]^{2} + \mathcal{V}_{eff}[\varphi(x,z)] \nn && + \frac{1}{2 \kappa} \Big( \mathcal{R}(x,z) - 2 \Lambda \Big) \Big\} + \mathcal{S}_{GHY} . \eqa
$\mathcal{G}^{MN}(x,z)$ given by the curly symbol is the metric tensor of the $(D+1)-$dimensional spacetime with its determinant inverse $\mathcal{G}(x,z)$, where $M$ and $N$ cover $0, ..., D$, including the extra-dimensional space. Accordingly, the curly $\mathcal{R}(x,z)$ is the $(D+1)$ dimensional Ricci scalar from the metric tensor $\mathcal{G}^{MN}(x,z)$. The last term $\mathcal{S}_{GHY}$ is the Gibbons-Hawking-York boundary action \cite{Gibbons_Hawking_York_I,Gibbons_Hawking_York_II}. This dual holographic effective bulk action enjoys the $(D+1)-$dimensional diffeomorphism invariance except for the boundary. Our strategy is to rewrite this diffeomorphism invariant bulk action in terms of the $D-$dimensional metric tensor with the lapse function and the shift vector field, referred to as the ADM formulation of the metric tensor \cite{ADM_Hamiltonian_Formulation}.

\section{Comparison between RG-derived emergent dual holography and string-based holographic duality}

The ADM formulation \cite{Holographic_RG_Review} is to decompose the $(D+1)-$dimensional metric tensor $[\bm{\mathcal{G}}]_{MN}(x,z)$ into the $D-$dimensional one $[\bm{g}]_{\mu\nu}(x,z)$ and the lapse function $\mathcal{N}(x,z)$ $\&$ the shift vector field $\mathcal{N}_{\mu}(x,z)$ as follows
\bqa && d s^{2} = \Big(\mathcal{N}^{2}(x,z) + \mathcal{N}_{\mu}(x,z) \mathcal{N}^{\mu}(x,z)\Big) d z^{2} + 2 \mathcal{N}_{\mu}(x,z) d z d x^{\mu} + g_{\mu\nu}(x,z) d x^{\mu} d x^{\nu} , \eqa
where
\bqa && [\bm{\mathcal{G}}]_{MN}(x,z) = \begin{pmatrix} \mathcal{N}^{2}(x,z) + \mathcal{N}_{\mu}(x,z) \mathcal{N}^{\mu}(x,z) & \mathcal{N}_{\mu}(x,z) \\ \mathcal{N}_{\mu}(x,z) & g_{\mu\nu}(x,z) \end{pmatrix} . \eqa
Accordingly, the $(D+1)-$dimensional Ricci scalar is decomposed as
\bqa && \mathcal{R}[\bm{\mathcal{G}}(x,z)] = \mathcal{R}[\bm{g}(x,z)] + R^{2}(x,z) - R_{\mu\nu}(x,z) R^{\mu\nu}(x,z) + \bm{\nabla}_{M} \zeta^{M}(x,z) . \eqa
Here, the induced Ricci tensor is given by the Lie derivative of the $(D+1)-$dimensional metric tensor along the direction of the lapse function, $R_{\mu\nu}(x,z) = \frac{1}{2} \Big(\mathcal{L}_{n(x,z)} \bm{\mathcal{G}}(x,z)\Big)_{\mu\nu}$ with $n^{M}(x,z) = \Big(\frac{1}{\mathcal{N}(x,z)}, - \frac{\mathcal{N}^{\mu}(x,z)}{\mathcal{N}(x,z)} \Big)$, where
\bqa && R_{\mu\nu}(x,z) = \frac{1}{2 \mathcal{N}(x,z)} \Big( \partial_{z} g_{\mu\nu}(x,z) - \mathcal{D}_{\mu} \mathcal{N}_{\nu}(x,z) - \mathcal{D}_{\nu} \mathcal{N}_{\mu}(x,z) \Big) . \eqa
The induced Ricci scalar is $R(x,z) = g^{\mu\nu}(x,z) R_{\mu\nu}(x,z)$. $\mathcal{D}_{\mu} \mathcal{N}_{\nu}(x,z) = \partial_{\mu} \mathcal{N}_{\nu}(x,z) - \Upsilon_{\mu\nu}^{\rho}[\bm{g}(x,z)] \mathcal{N}_{\rho}(x,z)$ is the $D-$dimensional covariant derivative, where $\Upsilon_{\mu\nu}^{\rho}[\bm{g}(x,z)]$ is the $D-$dimensional connection coefficient. The vector field $\zeta^{M}(x,z)$ is $\zeta^{M}(x,z) = - 2 R(x,z) n^{M}(x,z) + 2 n^{N}(x,z) \bm{\nabla}_{N} n^{M}(x,z)$ with $\bm{\nabla}_{M} \zeta^{M}(x,z) = \partial_{M} \zeta^{M}(x,z) + \Upsilon_{MN}^{M}(x,z) \zeta^{N}(x,z)$.

To clarify the connection between the string-theory based dual holographic action and the RG-derived effective bulk action, we perform the gauge fixing as follows
\bqa && \mathcal{N}(x,z) = 1 , ~~~~~ \mathcal{N}^{\mu}(x,z) = 0 . \eqa
As a result, we obtain
\bqa && \mathcal{S}_{eff}^{GR} = N \int_{0}^{z_{f}} d z \int d^{D} x \sqrt{g(x,z)} \Big\{ \frac{\mathcal{C}_{\varphi}}{2} [\partial_{z} \varphi(x,z)]^{2} + \frac{\mathcal{C}_{\varphi}}{2} g^{\mu\nu}(x,z) [\partial_{\mu} \varphi(x,z)] [\partial_{\nu} \varphi(x,z)] + \mathcal{V}_{eff}[\varphi(x,z)] \nn && + \mathcal{C}_{\xi} \Big(\mathcal{R}[\bm{g}(x,z)] + \frac{1}{4} [g^{\mu\nu}(x,z) \partial_{z} g_{\mu\nu}(x,z)]^{2} - \frac{1}{4} [\partial_{z} g^{\mu\nu}(x,z)] [\partial_{z} g_{\mu\nu}(x,z)] - \partial_{z} [g^{\mu\nu}(x,z) \partial_{z} g_{\mu\nu}(x,z)] \Big) [\varphi(x,z)]^{2} \nn && + \frac{1}{2 \kappa} \Big( \mathcal{R}[\bm{g}(x,z)] - 2 \Lambda + \frac{1}{4} [g^{\mu\nu}(x,z) \partial_{z} g_{\mu\nu}(x,z)]^{2} - \frac{1}{4} [\partial_{z} g^{\mu\nu}(x,z)] [\partial_{z} g_{\mu\nu}(x,z)] - \partial_{z} [g^{\mu\nu}(x,z) \partial_{z} g_{\mu\nu}(x,z)] \Big) \Big\} . \nn \label{Holography_GR_ADM} \eqa
We recall the emergent dual holographic bulk action
\bqa && \mathcal{S}_{eff}^{RG} = N \int_{0}^{z_{f}} d z \int d^{D} x \sqrt{g(x,z)} \Bigg\{ \frac{1}{2u} [\partial_{z} \varphi(x,z)]^{2} + \frac{\mathcal{C}_{\varphi}}{2} g^{\mu\nu}(x,z) [\partial_{\mu} \varphi(x,z)] [\partial_{\nu} \varphi(x,z)] + \mathcal{C}_{\xi} R(x,z) [\varphi(x,z)]^{2} \nn && + \mathcal{V}_{eff}[\varphi(x,z)] - \frac{1}{2 \lambda} \Big\{\partial_{z} g^{\mu\nu}(x,z) - g^{\mu\gamma}(x,z) \Big(\partial_{\gamma} \partial_{\gamma'} G_{xx'}[g_{\mu\nu}(x,z),\varphi(x,z)]\Big)_{x' \rightarrow x} g^{\gamma'\nu}(x,z)\Big\}^{2} + \frac{1}{2 \kappa} \Big( R(x,z) - 2 \Lambda \Big) \Bigg\} . \nn \label{Holography_RG_Bulk} \eqa

Comparing Eq. (\ref{Holography_RG_Bulk}) with Eq. (\ref{Holography_GR_ADM}), we observe that the main difference is the existence of \bqa && \beta_{g}^{\mu\nu}[g_{\mu\nu}(x,z), \varphi(x,z)] = g^{\mu\gamma}(x,z) \big(\partial_{\gamma} \partial_{\gamma'} G_{xx'}[g_{\mu\nu}(x,z),\varphi(x,z)]\big)_{x' \rightarrow x} g^{\gamma'\nu}(x,z) \label{RG_Flow_Metric_Green_Func} \eqa
in Eq. (\ref{Holography_RG_Bulk}) in addition to $\mathcal{C}_{\xi} \Big(\frac{1}{4} [g^{\mu\nu}(x,z) \partial_{z} g_{\mu\nu}(x,z)]^{2} - \frac{1}{4} [\partial_{z} g^{\mu\nu}(x,z)] [\partial_{z} g_{\mu\nu}(x,z)] - \partial_{z} [g^{\mu\nu}(x,z) \partial_{z} g_{\mu\nu}(x,z)] \Big) [\varphi(x,z)]^{2}$ of Eq. (\ref{Holography_GR_ADM}). This tensor-type RG $\beta-$function drives the evolution of the metric tensor along the extra-dimensional space, which allows the emergent geometry to be beyond the AdS type. To discuss this difference more precisely, we go to the ADM Hamiltonian formulation in the next section.

\section{Hamiltonian formulation}

To focus on the role of this $\beta-$function in the dual holographic description, we neglect the curvature-induced mass sector and the effective potential part for the dual scalar field and consider the following effective bulk action just for simplicity \cite{Holographic_RG_Review}
\bqa && \mathcal{S}_{eff}^{GR} = N \int_{0}^{z_{f}} d z \int d^{D} x \sqrt{g(x,z)} \mathcal{N}(x,z) \Big\{ \frac{\mathcal{C}_{\varphi}}{2 \mathcal{N}^{2}(x,z)} \Big(\partial_{z} \varphi(x,z) - \mathcal{N}^{\mu}(x,z) \partial_{\mu} \varphi(x,z)\Big)^{2} \nn && + \frac{\mathcal{C}_{\varphi}}{2} g^{\mu\nu}(x,z) [\partial_{\mu} \varphi(x,z)] [\partial_{\nu} \varphi(x,z)] + \frac{1}{2 \kappa} \Big( \mathcal{R}[\bm{g}(x,z)] - 2 \Lambda + R^{2}(x,z) - R_{\mu\nu}(x,z) R^{\mu\nu}(x,z) \Big) \Big\} . \eqa
Here, we introduced both the lapse function and the shift vector field to be responsible for the diffeomorphism invariance.

Following the ADM Hamiltonian formulation \cite{Holographic_RG_Review}, it is natural to consider two types of the canonical momenta as follows
\bqa && \pi^{\mu\nu}(x,z) = \frac{\delta \mathcal{S}_{eff}^{GR}}{\delta [\partial_{z} g_{\mu\nu}(x,z)]} = \frac{1}{\kappa} \sqrt{g(x,z)} \Big( R(x,z) g^{\mu\nu}(x,z) - R^{\mu\nu}(x,z) \Big) , \nn && \pi_{\varphi}(x,z) = \frac{\delta \mathcal{S}_{eff}^{GR}}{\delta [\partial_{z} \varphi(x,z)]} = \frac{\mathcal{C}_{\varphi}}{\mathcal{N}(x,z)} \sqrt{g(x,z)}\Big(\partial_{z} \varphi(x,z) - \mathcal{N}^{\mu}(x,z) \partial_{\mu} \varphi(x,z)\Big) . \label{Canonical_Momenta} \eqa
The former corresponds to the canonical momentum of the metric tensor and the latter does to that of the dual scalar field. Resorting to these canonical momenta, it is straightforward to perform the Legendre transformation and to obtain the effective Hamiltonian as follows
\bqa && \mathcal{S}_{eff}^{GR} = N \int_{0}^{z_{f}} d z \int d^{D} x \Big\{ \pi^{\mu\nu}(x,z) \partial_{z} g_{\mu\nu}(x,z) + \pi_{\varphi}(x,z) \partial_{z} \varphi(x,z) - \mathcal{N}(x,z) \mathcal{H}[\pi^{\mu\nu}(x,z), g^{\mu\nu}(x,z), \pi_{\varphi}(x,z), \varphi(x,z)] \nn && - \mathcal{N}_{\mu}(x,z) \mathcal{H}^{\mu}[\pi^{\mu\nu}(x,z), g^{\mu\nu}(x,z), \pi_{\varphi}(x,z), \varphi(x,z)] \Big\} . \eqa
Here, $\mathcal{H}[\pi^{\mu\nu}(x,z), g^{\mu\nu}(x,z), \pi_{\varphi}(x,z), \varphi(x,z)]$ coupled to the lapse function is given by
\bqa && \mathcal{H}[\pi^{\mu\nu}(x,z), g^{\mu\nu}(x,z), \pi_{\varphi}(x,z), \varphi(x,z)] \nn && = - \frac{\kappa}{2} \frac{1}{\sqrt{g(x,z)}} \Big( g_{\mu\rho}(x,z) g_{\nu\gamma}(x,z) - \frac{1}{D-1} g_{\mu\nu}(x,z) g_{\rho\gamma}(x,z) \Big) \pi^{\mu\nu}(x,z) \pi^{\rho\gamma}(x,z) + \frac{1}{2 \kappa} \sqrt{g(x,z)} \Big( \mathcal{R}[\bm{g}(x,z)] - 2 \Lambda \Big) \nn && - \frac{1}{2 \mathcal{C}_{\varphi}} \frac{1}{\sqrt{g(x,z)}} \pi_{\varphi}^{2}(x,z) + \frac{\mathcal{C}_{\varphi}}{2} \sqrt{g(x,z)} g^{\mu\nu}(x,z) [\partial_{\mu} \varphi(x,z)] [\partial_{\nu} \varphi(x,z)] , \eqa
rather expected. This effective Hamiltonian plays the role of the generator in the evolution along the extra-dimensional space as the ``time" translation operator. $\mathcal{H}^{\mu}[\pi^{\mu\nu}(x,z), g^{\mu\nu}(x,z), \pi_{\varphi}(x,z), \varphi(x,z)]$ coupled to the shift vector field is given by
\bqa && \mathcal{H}^{\mu}[\pi^{\mu\nu}(x,z), g^{\mu\nu}(x,z), \pi_{\varphi}(x,z), \varphi(x,z)] = - 2 \mathcal{D}_{\nu} \pi^{\mu\nu}(x,z) - \pi_{\varphi}(x,z) \partial^{\mu} \varphi(x,z) , \eqa
which gives rise to the $D-$dimensional diffeomorphism transformtion in a given $z$.

Since the effective bulk action depends on both the lapse function and the shift vector field only linearly, the path integral with respect to these fields results in two constraint equations, given by
\bqa && \mathcal{H}[\pi^{\mu\nu}(x,z), g^{\mu\nu}(x,z), \pi_{\varphi}(x,z), \varphi(x,z)] = 0 , ~~~~~ \mathcal{H}^{\mu}[\pi^{\mu\nu}(x,z), g^{\mu\nu}(x,z), \pi_{\varphi}(x,z), \varphi(x,z)] = 0 . \eqa
Physics of these two constraints are clarified, introducing the Hamilton's principal function $\mathcal{S}_{IR}[g^{\mu\nu}(x,z), \varphi(x,z)]$, which results in
\bqa && \pi^{\mu\nu}(x,z) = \frac{\delta \mathcal{S}_{IR}[g^{\mu\nu}(x,z), \varphi(x,z)]}{\delta g_{\mu\nu}(x,z)} , ~~~~~ \pi_{\varphi}(x,z) = \frac{\delta \mathcal{S}_{IR}[g^{\mu\nu}(x,z), \varphi(x,z)]}{\delta \varphi(x,z)} . \eqa
Combined with Eq. (\ref{Canonical_Momenta}), the RG flow of the metric tensor and that of the dual scalar field are expressed in terms of this principal function as follows
\bqa && \beta_{\mu\nu}^{g}[g_{\mu\nu}(x,z),\varphi(x,z)] \equiv \partial_{z} g_{\mu\nu}(x,z) \nn && = - \kappa \Big( g_{\mu\rho}(x,z) g_{\nu\gamma}(x,z) - \frac{1}{D-1} g_{\mu\nu}(x,z) g_{\rho\gamma}(x,z) \Big) \frac{1}{\sqrt{g(x,z)}} \frac{\delta \mathcal{S}_{IR}[g^{\mu\nu}(x,z), \varphi(x,z)]}{\delta g_{\rho\gamma}(x,z)} , \nn && \beta_{\varphi}[g_{\mu\nu}(x,z),\varphi(x,z)] \equiv \partial_{z} \varphi(x,z) = \frac{1}{\mathcal{C}_{\varphi}} \frac{1}{\sqrt{g(x,z)}} \frac{\delta \mathcal{S}_{IR}[g^{\mu\nu}(x,z), \varphi(x,z)]}{\delta \varphi(x,z)} . \label{RG_Flows_HJ} \eqa
These equations are nothing but the Hamilton's equations of motion for ``generalized coordinates" in the Hamilton-Jacobi formulation. These RG flows have to satisfy the Ward identity, given by the second constraint $\mathcal{H}^{\mu}[\pi^{\mu\nu}(x,z), g^{\mu\nu}(x,z), \pi_{\varphi}(x,z), \varphi(x,z)] = 0$ as
\bqa && 0 = - 2 \mathcal{D}_{\nu} \Bigg\{ \sqrt{g(x,z)} \Bigg(\frac{1}{\sqrt{g(x,z)}} \frac{\delta \mathcal{S}_{IR}[g^{\mu\nu}(x,z), \varphi(x,z)]}{\delta g_{\mu\nu}(x,z)}\Bigg) \Bigg\} \nn && - \sqrt{g(x,z)} \Bigg(\frac{1}{\sqrt{g(x,z)}} \frac{\delta \mathcal{S}_{IR}[g^{\mu\nu}(x,z), \varphi(x,z)]}{\delta \varphi(x,z)}\Bigg) \partial^{\mu} \varphi(x,z) . \label{Ward_Identity_HJ} \eqa
Finally, the Hamilton's principal function is determined by the first constraint $\mathcal{H}[\pi^{\mu\nu}(x,z), g^{\mu\nu}(x,z), \pi_{\varphi}(x,z), \varphi(x,z)] = 0$, given by
\bqa && 0 = - \frac{\kappa}{2} \sqrt{g(x,z)} \Big( g_{\mu\rho}(x,z) g_{\nu\gamma}(x,z) - \frac{1}{D-1} g_{\mu\nu}(x,z) g_{\rho\gamma}(x,z) \Big) \nn && \times \Bigg(\frac{1}{\sqrt{g(x,z)}} \frac{\delta \mathcal{S}_{IR}[g^{\mu\nu}(x,z), \varphi(x,z)]}{\delta g_{\mu\nu}(x,z)}\Bigg) \Bigg(\frac{1}{\sqrt{g(x,z)}} \frac{\delta \mathcal{S}_{IR}[g^{\mu\nu}(x,z), \varphi(x,z)]}{\delta g_{\rho\gamma}(x,z)}\Bigg) + \frac{1}{2 \kappa} \sqrt{g(x,z)} \Big( \mathcal{R}[\bm{g}(x,z)] - 2 \Lambda \Big) \nn && - \frac{1}{2 \mathcal{C}_{\varphi}} \sqrt{g(x,z)} \Bigg(\frac{1}{\sqrt{g(x,z)}} \frac{\delta \mathcal{S}_{IR}[g^{\mu\nu}(x,z), \varphi(x,z)]}{\delta \varphi(x,z)}\Bigg)^{2} + \frac{\mathcal{C}_{\varphi}}{2} \sqrt{g(x,z)} g^{\mu\nu}(x,z) [\partial_{\mu} \varphi(x,z)] [\partial_{\nu} \varphi(x,z)] , \label{HJ_String_Holography} \eqa
refereed to as the Hamilton-Jacobi equation.

Actually, this procedure to find the effective on-shell action has been well known in the context of holographic renormalization \cite{Holographic_Duality_IV,Holographic_Duality_V,Holographic_Duality_VI}. The only novel ingredient that we claim is that the Hamilton's principal function, more physically, the effective on-shell action can be identified with the effective IR boundary action of the RG-based emergent dual holographic description, given by
\bqa && \mathcal{S}_{IR}[g^{\mu\nu}(x,z), \varphi(x,z)] \nn && = - \ln \int D \phi_{\alpha}(x) \exp\Big[ - \int d^{D} x \sqrt{g(x,z)} \Big\{ g^{\mu\nu}(x,z) [\partial_{\mu} \phi_{\alpha}(x)] [\partial_{\nu} \phi_{\alpha}(x)] + [m^{2} - i \varphi(x,z)] \phi_{\alpha}^{2}(x) \Big\} \Big] . \label{IR_Boundary_Action} \eqa
This IR boundary action serves as the IR boundary condition in the RG-based emergent dual holographic description. Recently, we derived the Hamilton-Jacobi equation from Eq. (\ref{Holography_RG_Bulk}), which can be regarded as the inverse procedure of the Wilsonian RG transformation, given by \cite{RG_GR_Geometry_I_Kim}
\bqa && \frac{1}{\sqrt{g(x,z)}} \frac{\partial}{\partial z} \mathcal{I}[g_{\mu\nu}(x,z), \varphi(x,z)] = \frac{\lambda}{2} \Big\{ \frac{1}{\sqrt{g(x,z)}} \frac{\partial \mathcal{I}[g_{\mu\nu}(x,z), \varphi(x,z)]}{\partial g^{\mu\nu}(x,z)} \Big\}^{2} \nn && + \beta_{g}^{\mu\nu}[g_{\mu\nu}(x,z), \varphi(x,z)] \Big\{ \frac{1}{\sqrt{g(x,z)}} \frac{\partial \mathcal{I}[g_{\mu\nu}(x,z), \varphi(x,z)]}{\partial g^{\mu\nu}(x,z)} \Big\} + \frac{1}{2 \kappa} \Big( R(x,z) - 2 \Lambda \Big) \nn && - \frac{u}{2} \Big\{ \frac{1}{\sqrt{g(x,z)}} \frac{\partial \mathcal{I}[\varphi(x,z), g_{\mu\nu}(x,z)]}{\partial \varphi(x,z)} \Big\}^{2} + \frac{\mathcal{C}_{\varphi}}{2} g^{\mu\nu}(x,z) [\partial_{\mu} \varphi(x,z)] [\partial_{\nu} \varphi(x,z)] . \label{HJ_RG_Holography} \eqa
Here, $\mathcal{I}[g_{\mu\nu}(x,z), \varphi(x,z)]$ is the effective on-shell action, which has to be the IR boundary action $\mathcal{S}_{IR}[g^{\mu\nu}(x,z), \varphi(x,z)]$ for self-consistency of the framework. Comparing Eq. (\ref{HJ_RG_Holography}) with Eq. (\ref{HJ_String_Holography}), we realize that the main difference is the existence of the RG $\beta-$function in Eq. (\ref{HJ_RG_Holography}). Of course, there are tensor indices in the gravity sector of Eq. (\ref{HJ_String_Holography}), the absence of which would imply that the diffeomorphism invariance may not be carefully implemented in Eq. (\ref{HJ_RG_Holography}) during the Wilsonian RG-transformation procedure. Moreover, it turns out that $\mathcal{I}[g_{\mu\nu}(x,z), \varphi(x,z)]$ does not depend on the coordinate of the extra-dimensional space explicitly, which results in $\frac{\partial}{\partial z} \mathcal{I}[g_{\mu\nu}(x,z), \varphi(x,z)] = 0$ in Eq. (\ref{HJ_RG_Holography}).

The above discussion suggests that the effective on-shell action as the solution of the Hamilton-Jacobi equation (\ref{HJ_String_Holography}) is the IR effective action Eq. (\ref{IR_Boundary_Action}), which serves as the IR boundary condition in the RG-based dual holographic description, if the information of the RG flows given by Eq. (\ref{RG_Flows_HJ}) is introduced into the Hamilton-Jacobi equation (\ref{HJ_String_Holography}) as that in Eq. (\ref{HJ_RG_Holography}). It is natural to replace the RG $\beta-$function of the metric tensor in Eq. (\ref{RG_Flows_HJ}) with the Ricci tensor as follows
\bqa && \beta_{\mu\nu}^{g}[g_{\mu\nu}(x,z),\varphi(x,z)] = - R_{\mu\nu}(x,z) . \label{Ricci_Flow} \eqa
Then, the RG flow of the metric tensor is given by the Ricci flow $\partial_{z} g_{\mu\nu}(x,z) = - R_{\mu\nu}(x,z)$ \cite{Ricci_Flow_0,Ricci_Flow_I,Ricci_Flow_II,Ricci_Flow_III,Ricci_Flow_IV,Ricci_Flow_V}. Actually, it has been shown that the RG flow of the metric tensor in the holographic RG formulation gives rise to the Ricci flow \cite{Holographic_RG_Flow_Ricci_Flow_I,Holographic_RG_Flow_Ricci_Flow_II}. Motivated from this conjecture, we replaced the RG $\beta-$function of Eq. (\ref{RG_Flow_Metric_Green_Func}) with the Ricci tensor \cite{RG_GR_Geometry_I_Kim}. Here, the gradient expansion with respect to the mass parameter in the Green's function would allow other terms involved with derivatives of dual scalar fields, but they are all neglected for simplicity, where the diffeomorphism invariance is not taken into account carefully. In appendix, we give two supporting arguments on this Ricci flow conjecture for the RG $\beta-$function. As a result, we proposed the following effective bulk action
\bqa && \mathcal{S}_{eff}^{RG} = N \int_{0}^{z_{f}} d z \int d^{D} x \sqrt{g(x,z)} \Big\{ \frac{1}{2u} [\partial_{z} \varphi(x,z)]^{2} + \frac{\mathcal{C}_{\varphi}}{2} g^{\mu\nu}(x,z) [\partial_{\mu} \varphi(x,z)] [\partial_{\nu} \varphi(x,z)] + \mathcal{V}_{eff}[\varphi(x,z)] \nn && - \frac{1}{2 \lambda} \Big(\partial_{z} g^{\mu\nu}(x,z) + R^{\mu\nu}(x,z)\Big) \Big(\partial_{z} g_{\mu\nu}(x,z) + R_{\mu\nu}(x,z)\Big) + \frac{1}{2 \kappa} \Big( R(x,z) - 2 \Lambda \Big) \Big\} , \label{RG_Ricci_Flow_Bulk_Action} \eqa
where the RG flow of the metric tensor has been manifestly realized \cite{RG_GR_Geometry_I_Kim}.

Considering the diffeomorphism invariance, finally we propose
\bqa && Z[\varphi_{ext}(x)] = \int D \phi_{\alpha}(x) D \varphi(x,z) D \mathcal{G}_{MN}(x,z) ~ \delta\Big(g^{\mu\nu}(x,0) - \delta^{\mu\nu}\Big) \delta\Big(\varphi(x,0) - \varphi_{ext}(x)\Big) \nn && \exp\Big[ - \int d^{D} x \sqrt{g(x,z_{f})} \Big\{ g^{\mu\nu}(x,z_{f}) [\partial_{\mu} \phi_{\alpha}(x)] [\partial_{\nu} \phi_{\alpha}(x)] + [m^{2} - i \varphi(x,z_{f})] \phi_{\alpha}^{2}(x) + \xi R(x,z_{f}) \phi_{\alpha}^{2}(x) \Big\} \nn && - N \int_{0}^{z_{f}} d z \int d^{D} x \sqrt{\mathcal{G}(x,z)} \Big\{ \frac{\mathcal{C}_{\varphi}}{2} \mathcal{G}^{MN}(x,z) [\partial_{M} \varphi(x,z)] [\partial_{N} \varphi(x,z)] + \mathcal{C}_{\xi} \mathcal{R}(x,z) [\varphi(x,z)]^{2} + \mathcal{V}_{eff}[\varphi(x,z)] \nn && + \frac{1}{2 \kappa} \Big( \mathcal{R}(x,z) - 2 \Lambda \Big) - \frac{\mathcal{C}_{1}}{2} \mathcal{R}_{MN}(x,z) \mathcal{R}^{MN}(x,z) + \frac{\mathcal{C}_{2}}{2} \mathcal{R}^{2}(x,z) \Big\} - \mathcal{S}_{TOP} - \mathcal{S}_{GHY} \Big] , \label{Holography} \eqa
where $\mathcal{S}_{TOP}$ is the topological invariant from higher-derivative curvature terms \cite{Higher_Derivative_Curvature} and $\mathcal{S}_{GHY}$ is the Gibbons-Hawking-York boundary action. There are two main updated ingredients in this holographic dual effective field theory. One crucial point is the introduction of the IR boundary action into the string-theory based conventional dual holographic framework, which completes the deep connection between UV microscopic and IR macroscopic degrees of freedom. The other is the introduction of higher-derivative curvature terms \cite{Higher_Derivative_Curvature}, which gives rise to the RG flow of the metric tensor.

\section{Conclusion}

The main motivation of the present study is to propose how to overcome two kinds of unsatisfactory points in the RG-derived dual holographic effective field theory: (i) This effective field theory is derived approximately, where the recursive RG transformation in the path integral formulation is performed in the linearized fashion, and (ii) $(D+1)-$dimensional diffeomorphism invariance is explicitly broken, where $D$ is the spacetime dimension of the UV quantum field theory. In particular, the Ward identity is not manifested in the RG-derived dual holography as a result of (ii), which implies that we cannot justify the resulting dual holographic effective field theory based on the Ward identity. To resolve these weak aspects, we consider to compare the RG-derived dual holographic effective field theory with the string-theory based conventionally utilized Einstein-scalar dual holographic field theory, where the $(D+1)-$dimensional diffeomorphism invariance is explicit except for the boundary conditions. Of course, there is an essential problem in the string-theory based dual holography for general applications to condensed matter physics: It is not clear at all how UV microscopic degrees of freedom are transformed into IR emergent macroscopic fields. In this study, we tried to solve these problems of both dual holographic field theories based on the Hamilton-Jacobi formulation.

First, we reformulate the string-theory based dual holographic effective field theory in the ADM representation, which decomposes the $(D+1)-$dimensional metric tensor into the $D-$dimensional metric tensor with the lapse function and shift vector field. As a result,
the evolution of the metric tensor along the extra-dimensional space becomes manifested in the string-theory based holographic dual field theory.
To compare this effective field theory with the RG-derived holographic dual field theory, we introduce the gauge-fixing condition for both lapse function and shift vector fields, and obtain Eq. (\ref{Holography_GR_ADM}). Compared to the RG-derived holographic dual field theory Eq. (\ref{Holography_RG_Bulk}), it becomes transparent that (i) the tensorial index structure is not carefully taken into account in the RG-derived holographic dual field theory Eq. (\ref{Holography_RG_Bulk}) and (ii) the genuine RG flow of the metric tensor along the extra-dimensional space is not introduced into the string-theory based dual holographic effective field theory Eq. (\ref{Holography_GR_ADM}). The unsatisfactory point (i) is purely a fault in the RG-transformation implementation for $T_{\mu\nu} T^{\mu\nu}$ effective interactions, but is easily corrected through keeping such tensorial indices more carefully. On the other hand, the lack of the RG $\beta-$function contribution in the string-theory based dual holographic effective field theory Eq. (\ref{Holography_GR_ADM}) shows the fundamental difficulty in the application of the string-theory based dual holography into the condensed matter physics. We point out that the vanishing $\beta-$function contribution in the string-theory based dual holographic effective field theory Eq. (\ref{Holography_GR_ADM}) is consistent with the fact that only AdS geometry is considered in the string-theory based dual holography.

To clarify this difference between Eqs. (\ref{Holography_GR_ADM}) and (\ref{Holography_RG_Bulk}) and to overcome the unsatisfactory point in the string-theory based dual holographic effective field theory Eq. (\ref{Holography_GR_ADM}), second, we consider the ADM Hamiltonian gravity formulation of the string-theory based dual holographic effective field theory Eq. (\ref{Holography_GR_ADM}). Following the standard well-known procedure, we obtain the Hamilton-Jacobi equation (\ref{HJ_String_Holography}) for an effective on-shell action of Eq. (\ref{Holography_GR_ADM}) with the Ward identity Eq. (\ref{Ward_Identity_HJ}) to describe the energy-momentum tensor current conservation law, where the RG-flow equations (\ref{RG_Flows_HJ}) are nothing but the Hamilton's equations of motion for ``generalized coordinates". Here, the essential point of the present study appears. We claim that the IR boundary effective action Eq. (\ref{IR_Boundary_Action}) of the RG-derived holographic dual effective field theory can be the on-shell action solution of the Hamilton-Jacobi equation (\ref{HJ_String_Holography}) of the string-theory based dual holographic effective field theory Eq. (\ref{Holography_GR_ADM}). To confirm this claim, we show the Hamilton-Jacobi equation (\ref{HJ_RG_Holography}) of the RG-derived holographic dual field theory Eq. (\ref{Holography_RG_Bulk}), which has been discussed in our previous work \cite{RG_GR_Geometry_I_Kim}. We recall that the IR boundary effective action Eq. (\ref{IR_Boundary_Action}) must be the solution of the Hamilton-Jacobi equation (\ref{HJ_RG_Holography}) of the RG-derived holographic dual field theory Eq. (\ref{Holography_RG_Bulk}). Comparing the RG-based Hamilton-Jacobi Eq. (\ref{HJ_String_Holography}) with the string-theory based Hamilton-Jacobi Eq. (\ref{HJ_RG_Holography}), we observe that the essential difference between Eqs. (\ref{HJ_String_Holography}) and (\ref{HJ_RG_Holography}) is the absence [Eq. (\ref{HJ_String_Holography})] or presence [Eq. (\ref{HJ_RG_Holography})] of the $\beta-$function, which gives rise to the RG flow of the metric tensor. Of course, there is additional minor difference in the tensorial index structure as pointed out earlier.

To make the IR boundary effective action Eq. (\ref{IR_Boundary_Action}) of the RG-derived holographic dual effective field theory be the on-shell action solution of the Hamilton-Jacobi equation (\ref{HJ_String_Holography}) of the string-theory based dual holographic effective field theory Eq. (\ref{Holography_GR_ADM}), finally, we introduce the $\beta-$function contribution into the string-theory based dual holographic effective field theory Eq. (\ref{Holography_GR_ADM}) as the form of higher derivative curvature terms. As a result, we propose an effective holographic dual field theory Eq. (\ref{Holography}), where (i) the IR boundary effective action is introduced into the bulk holographic dual field theory, which serves as the deep connection between UV microscopic degrees of freedom and IR emergent macroscopic field variables through the IR boundary condition, and (ii) higher-derivative curvature terms are taken into account in the bulk effective action with diffeomorphism invariance, which gives rise to the RG flow from UV to IR, expected to be applicable to the geometry beyond the AdS type.

The remaining inevitable task is to confirm that the proposed holographic dual effective field theory Eq. (\ref{Holography}) is internally consistent. We can argue the internal consistency of Eq. (\ref{Holography}) in two ways. First of all, we have to show that the IR boundary action in Eq. (\ref{Holography}) is the on-shell effective-action solution of the Hamilton-Jacobi equation, derived from the bulk effective action of Eq. (\ref{Holography}) based on the ADM Hamiltonian formulation. This point is currently under investigation. Second, we can show that taking the limit of $z_{f} \rightarrow 0$ in Eq. (\ref{Holography}), the resulting effective field theory reproduces the one-loop RG-transformation result of the UV quantum field theory. Although we do not show our explicit demonstration here, we find that Eq. (\ref{Holography}) is essentially reduced into the one-loop RG-transformation result of the UV quantum field theory, indeed, in the $z_{f} \rightarrow 0$ limit, where some modifications arise due to both higher-derivative curvature terms and diffeomorphism invariance.

%
%The main defect of the holographic dual effective field theory Eq. (\ref{RG_Holography_EFT}) is the lack of the diffeomorphism invariance. On the other hand, the string-theory based dual holography framework does not show any clear connections between UV microscopic and IR emergent degrees of freedom. In this study, we compared these two frameworks based on the Hamilton-Jacobi formulation and suggested corrections to the string-theory based dual holographic effective field theory. As a result, we introduced (i) the IR boundary effective action into the bulk holographic dual field theory, which serves as the deep connection between UV microscopic degrees of freedom and IR emergent field variables through the IR boundary condition, and (ii) higher-derivative curvature terms into the bulk effective action with diffeomorphism invariance, which gives rise to the RG flow from UV to IR, expected to be applicable to the geometry beyond the AdS type. All these ingredients are summarized in Eq. (\ref{Holography}).
%

\section*{Acknowledgement}

K.-S. Kim was supported by the Ministry of Education, Science, and Technology (No. 2011-0030046) of the National Research Foundation of Korea (NRF) and by TJ Park Science Fellowship of the POSCO TJ Park Foundation. KSK appreciates helpful discussions with Shinsei Ryu and Chanyong Park.

\section*{Appendix}

First, we justify Eq. (\ref{Ricci_Flow}) physically. We recall the RG-flow equation of the metric tensor, given by
\bqa
&& \beta_{\mu\nu}^{g}[g_{\mu\nu}(x,z),\varphi(x,z)] \equiv \partial_{z} g_{\mu\nu}(x,z) \nn && = - \kappa \Big( g_{\mu\rho}(x,z) g_{\nu\gamma}(x,z) - \frac{1}{D-1} g_{\mu\nu}(x,z) g_{\rho\gamma}(x,z) \Big) \frac{1}{\sqrt{g(x,z)}} \frac{\delta \mathcal{S}_{IR}[g^{\mu\nu}(x,z), \varphi(x,z)]}{\delta g_{\rho\gamma}(x,z)} , \nonumber
\eqa
which is nothing but the Hamilton's equation of motion for the generalized coordinate $g_{\mu\nu}(x,z)$. Considering that the energy-momentum tensor is given by
\bqa
&& T^{\rho\gamma}(x,z) = - \frac{2}{\sqrt{g(x,z)}} \frac{\delta \mathcal{S}_{IR}[g^{\mu\nu}(x,z), \varphi(x,z)]}{\delta g_{\rho\gamma}(x,z)} , \nonumber
\eqa
we obtain
\bqa
&& \beta_{\mu\nu}^{g}[g_{\mu\nu}(x,z),\varphi(x,z)] = \frac{\kappa}{2} \Big( g_{\mu\rho}(x,z) g_{\nu\gamma}(x,z) - \frac{1}{D-1} g_{\mu\nu}(x,z) g_{\rho\gamma}(x,z) \Big) T^{\rho\gamma}(x,z) . \nonumber
\eqa
Actually, the $\beta-$function for the RG flow of the metric tensor, given by
\bqa
&& \beta_{g}^{\mu\nu}[g_{\mu\nu}(x,z), \varphi(x,z)] = g^{\mu\gamma}(x,z) \big(\partial_{\gamma} \partial_{\gamma'} G_{xx'}[g_{\mu\nu}(x,z),\varphi(x,z)]\big)_{x' \rightarrow x} g^{\gamma'\nu}(x,z) \nonumber
\eqa
in the RG-based holographic dual field theory, is consistent with the above equation $\beta_{\mu\nu}^{g}[g_{\mu\nu}(x,z),\varphi(x,z)] \propto \Big\langle T_{\mu\nu}(x,z) \Big\rangle$. Resorting to the Einstein equation,
\bqa
&& R_{\mu\nu}(x,z) - \frac{1}{2} R(x,z) g_{\mu\nu} + \Lambda g_{\mu\nu} = \kappa T_{\mu\nu} , \nonumber
\eqa
we can rewrite the above expression as follows
\bqa
\beta_{\mu\nu}^{g}[g_{\mu\nu}(x,z),\varphi(x,z)] &=& \frac{1}{2} \Big( g_{\mu\rho}(x,z) g_{\nu\gamma}(x,z) - \frac{1}{D-1} g_{\mu\nu}(x,z) g_{\rho\gamma}(x,z) \Big) \Big( R^{\rho\gamma}(x,z) - \frac{1}{2} R(x,z) g^{\rho\gamma} + \Lambda g^{\rho\gamma} \Big) . \nonumber
\eqa
This is our physical explanation for Eq. (\ref{Ricci_Flow}).

Second, we consider our brute-force calculation of the Green's function on a general curved spacetime background. Here, we follow Ref. \cite{Entanglement_Entropy_Heat_Kernel}, a part of which is summarized in Ref. \cite{Einstein_Klein_Gordon_RG_Kim}. We recall that the Green's function is given by
\begin{align}
  & \Big\{- \frac{1}{\sqrt{g(x,z)}} \partial_{\mu} \Big( \sqrt{g(x,z)} g^{\mu\nu}(x,z) \partial_{\nu} \Big) + \frac{m^{2}}{\epsilon} \Big\} G[x,x';g_{\mu\nu}(x,z)] = \frac{1}{\sqrt{g(x,z)}} \delta^{(D)}(x-x') , \nonumber
\end{align}
where the dual order-parameter field has been neglected just for simplicity. One can reformulate this Green's function in terms of the geometric information such as metric and curvature, considering the heat kernel expansion. Introducing the heat kernel $K(s,x,x';z) \equiv \langle x z | e^{- s \mathcal{D}} | x' z \rangle$ with the differential operator $\mathcal{D} \equiv - \frac{1}{\sqrt{g(x,z)}} \partial_{\mu} \Big( \sqrt{g(x,z)} g^{\mu\nu}(x,z) \partial_{\nu} \Big) + \frac{m^{2}}{\epsilon}$, which satisfies the diffusion equation $(\partial_{s} + \mathcal{D}) K(s,x,x';z) = 0$ with an initial condition $K(s = 0, x, x';z) = \frac{1}{\sqrt{g(x,z)}} \delta^{(D)}(x-x')$, we obtain
\bqa && G(x,x';z) = \int_{\varepsilon^{2}}^{\infty} d s ~ K(s,x,x';z) . \nonumber \eqa
Here, $\varepsilon$ is a UV cutoff to cure the UV divergence of the entanglement entropy, for example.

One may take an expansion for small $s$ in the heat kernel, given by
\bqa && K(s,x,x' \rightarrow x;z) = \frac{1}{(4\pi s)^{\frac{d}{2}}} \sum_{n = 0} a_{n}(x,z) s^{n} . \nonumber \eqa
These bulk coefficients $a_{n}(x,z)$ are given by
\begin{align}
   a_{0}(x,z) &= 1 ,
    \nonumber \\
  a_{1}(x,z) &= \frac{1}{6} R(x,z) - \frac{m^{2}}{\varepsilon} ,
                \nonumber \\
 a_{2}(x,z) &= \frac{1}{180} R^{2}_{\mu\nu\alpha\beta}(x,z) - \frac{1}{180} R_{\mu\nu}^{2}(x,z)
                               \nonumber \\
&+ \frac{1}{6} \frac{1}{\sqrt{g(x,z)}} \partial_{\mu} \Big( \sqrt{g(x,z)} g^{\mu\nu}(x,z) \partial_{\nu} \Big) \Big( \frac{1}{5} R(x,z) - m^{2} \Big) + \frac{1}{2} \Big( \frac{1}{6} R(x,z) - \frac{m^{2}}{\varepsilon} \Big)^{2} \nonumber
\end{align}
for general curvature tensors, well discussed in Ref. \cite{Entanglement_Entropy_Heat_Kernel}. As a result, we express the Green's function in terms of the metric and curvature at $z$ as follows
\bqa
&& G(x,x;z) = \int_{\varepsilon^{2}}^{\infty} d s ~ K(s,x,x;z) = \int_{\varepsilon^{2}}^{\infty} d s
~ \sum_{n = 0}^{\infty} a_{n}(x,z) s^{\frac{n-D}{2}}
\nn &&\quad = \int_{\varepsilon^{2}}^{\infty} d s ~ \Big\{ a_{0}(x,z) s^{-\frac{D}{2}} +
a_{2}(x,z) s^{\frac{2 - D}{2}} + a_{4}(x,z) s^{\frac{4 - D}{2}} +
\ldots
\Big\}
\nn &&\quad = \int_{\varepsilon^{2}}^{\infty} d s ~ \frac{1}{(4 \pi)^{\frac{D}{2}}} \Big[
s^{-\frac{D}{2}}
+ \Big( \frac{1}{6} R(x,z) - \frac{m^{2}}{\epsilon} \Big) s^{\frac{2 - D}{2}}
+ \Big\{ \frac{1}{180}
R^{2}_{\mu\nu\alpha\beta}(x,z) - \frac{1}{180} R_{\mu\nu}^{2}(x,z)
\nn &&\quad + \frac{1}{6} \frac{1}{\sqrt{g(x,z)}} \partial_{\mu} \Big(
\sqrt{g(x,z_{f})} g^{\mu\nu}(x,z_{f}) \partial_{\nu} \Big) \Big( \frac{1}{5}
R(x,z) - \frac{m^{2}}{\epsilon} \Big)
+ \frac{1}{2} \Big( \frac{1}{6} R(x,z) - \frac{m^{2}}{\epsilon} \Big)^{2} \Big\} s^{\frac{4 - D}{2}} +
\ldots \Big] . \nonumber \eqa
Applying the $x-$derivative twice, one can find the $\beta-$function for the RG flow of the metric tensor. Here, the gradient expansion with respect to the mass parameter in the Green's function would allow other terms involved with derivatives of dual scalar fields when introduced, but they are all neglected for simplicity. As a result, one can obtain Eq. (\ref{RG_Ricci_Flow_Bulk_Action}) for an effective holographic dual field theory as an approximate expression of Eq. (\ref{Holography_RG_Bulk}), which is an actual RG-derived effective bulk action.


\begin{thebibliography}{9}
\bibitem{Comment_Higher_Spin_GT} One may suggest a higher-spin gauge theory \cite{Higher_Spin_Gauge_Theory_I,Higher_Spin_Gauge_Theory_II,Higher_Spin_Gauge_Theory_III,Higher_Spin_Gauge_Theory_IV} as an effective holographic dual field theory. However, it turns out that the nonlocal diffeomorphism invariance, responsible for the emergence of the higher-spin gauge theory \cite{Holography_Higher_Spin_RG_I,Holography_Higher_Spin_RG_II,Holography_Higher_Spin_RG_III}, is explicitly broken due to the existence of effective interactions \cite{Einstein_Klein_Gordon_RG_Kim}, although the nonlocal diffeomorphism invariance exists in the absence of such interactions. In this case, the ultimate fate of higher spin fields has not been clarified as far as we know. See Ref. \cite{Einstein_Klein_Gordon_RG_Kim} for more discussions.
\end{thebibliography}
\end{document}